\documentclass[twocolumn,times]{aastex62}
\usepackage{mathtools}

\usepackage{graphicx}
\usepackage{lipsum}

\begin{document}

\title{A Full Resolution of the 450\,$\micron$ Extragalactic Background Light}

\correspondingauthor{Qi-Ning Hsu, Chian-Chou Chen}
\email{qnhsu@asiaa.sinica.edu.tw, ccchen@asiaa.sinica.edu.tw}

\author[0000-0002-3313-8001]{Qi-Ning Hsu}
\affiliation{Graduate Institute of Astronomy, National Tsing Hua University, No. 101, Section 2, Kuang-Fu Road, Hsinchu 30013, Taiwan}
\affiliation{Academia Sinica Institute of Astronomy and Astrophysics (ASIAA), No. 1, Section 4, Roosevelt Rd., Taipei 10617, Taiwan}

\author[0000-0002-6319-1575]{L. L. Cowie}
\affiliation{Institute for Astronomy, University of Hawaii, 2680 Woodlawn Drive, Honolulu, HI 96822, USA}

\author[0000-0002-3805-0789]{Chian-Chou Chen}
\affiliation{Academia Sinica Institute of Astronomy and Astrophysics (ASIAA), No. 1, Section 4, Roosevelt Rd., Taipei 10617, Taiwan}

\author[0000-0002-3306-1606]{A. J. Barger}
\affiliation{Department of Astronomy, University of Wisconsin-Madison, 475 N. Charter Street, Madison, WI 53706, USA}
\affiliation{Department of Physics and Astronomy, University of Hawaii, 2505 Correa Road, Honolulu, HI 96822, USA}
\affiliation{Institute for Astronomy, University of Hawaii, 2680 Woodlawn Drive, Honolulu, HI 96822, USA}

\begin{abstract}

The extragalactic background light (EBL) is the cumulative radiation outside the Milky Way. The determination of its corresponding primary emitting sources as well as its total energy level across the entire electromagnetic spectrum has profound implications for both cosmology and galaxy formation.
However, the detailed origin of the EBL at far-infrared wavelengths, particularly those close to the peak of the cosmic infrared background, remains unclear. Here we report the results of our ongoing SCUBA-2 450\,$\micron$ survey of 10 massive galaxy cluster fields.
By exploiting the strong gravitational lensing offered by these clusters, we obtain significant counts down to an unprecedented depth of $\sim$0.1\,mJy at this wavelength, about ten times deeper than that reached by any other previous survey. The cumulative energy density based on the counts is 138.1$^{+23.9}_{-19.3}$ Jy degree$^{-2}$, or 0.45$^{+0.08}_{-0.06}$ MJy sr$^{-1}$. Comparing our measurements to those made by the COBE and Planck satellites, we find that at this flux density level, the 450\,$\micron$ EBL is entirely resolved by our SCUBA-2 observations. Thus, we find for the first time that discrete sources produce fully to the 450\,$\micron$ EBL, and that about half of it comes from sources with sub-mJy flux densities.
Our deep number counts provide strong constraints on galaxy formation models.

\end{abstract}

\keywords{galaxies: clusters --- galaxies: number counts --- EBL --- submillimeter galaxies}

\section{Introduction} \label{sec:intro}

The cosmic energy budget, encompassing the total amount of energy radiated throughout the Universe, is a fundamental aspect of modern astrophysics. Central to this budget is the extragalactic background light (EBL), which represents the integrated emission across the electromagnetic spectrum from astrophysical sources outside the Milky Way (see, e.g., the review by \citealt{2016RSOS....350555C}). Observational measurements of the EBL at different wavelengths allow one to understand its energy distribution, and, in principle, can provide insights on the dominant contributors to the EBL (see, e.g., the discussion in \citealt{2018ApSpe..72..663H}).

One common way to put constraints on the EBL is to perform imaging surveys using ground-based facilities, which allow one to construct source number counts and calculate the integrated energy densities (e.g., the early results in the far-infrared (FIR)/submillimeter by \citealt{1997ApJ...490L...5S, 1999ASPC..191..279B, 2002AJ....123.2197C}).
By comparing with satellite measurements having lower spatial resolution, which, in principle, take into account any diffuse emission, one can estimate the energy contributions to the EBL from galaxies and understand whether galaxies are the dominant contributors.
Surveys have found that the optical and near-infrared EBL primarily originate from the directly observed star formation in galaxies, while the FIR/submillimeter EBL mainly comes from the thermal emission from interstellar dust reradiated starlight \citep{1996A&A...308L...5P,1998ApJ...508..123F,2006A&A...451..417D}. These dusty galaxies, sometimes called submillimeter galaxies (SMGs), are typically characterized by intense dust emission indicating high rates of star formation (see, e.g., review articles by \citealt{2002PhR...369..111B, 2014PhR...541...45C}).

The advent of space-based telescopes like the Herschel Space Observatory and ground-based facilities like the Atacama Large Millimeter/submillimeter Array (ALMA) has revolutionized our ability to perform either wide-field or deep FIR surveys, enabling unprecedented studies of galaxy number counts \citep{2010A&A...518L..21O, 2011A&A...532A..49B, 2022A&A...658A..43G, 2023ApJ...952...28C, 2023MNRAS.518.1378C}. 
These observations, coupled with theoretical models, provide a comprehensive view of the FIR Universe and its connection to other astrophysical phenomena (e.g., \citealt{2016MNRAS.462.3854L, 2019MNRAS.489.4196L, 2021MNRAS.502.2922H}). 
However, due to the confusion limits of Herschel and the inefficient survey capability of ALMA with its small field-of-view, the number counts at wavelengths close to the peak of the FIR EBL ($\sim200-300$\,\micron; e.g., \citealt{2019ApJ...877...40O}) remain limited to the brightest end (e.g., \citealt{2010A&A...518L..21O}).

SCUBA-2, a state-of-the-art submillimeter camera, offers unmatched sensitivity and high angular resolution at 450\,$\micron$. Its optimized design for deep submillimeter observations makes it an excellent tool for studying dust-rich galaxies \citep{2013MNRAS.430.2513H}. By harnessing the sensitivity of SCUBA-2 at submillimeter wavelengths, we can detect and characterize faint sources at 450\,$\micron$, enabling comprehensive studies of galaxy number counts close to the peak of the FIR EBL and a deeper exploration of the submillimeter Universe (\citealt{2013MNRAS.432...53G, 2013MNRAS.436.1919C, 2017MNRAS.464.3369Z, 2017ApJ...850...37W, 2020ApJ...889...80L, 2022ApJ...934...56B, Gao2024aa}). 

Additionally, the technique of gravitational lensing provides a unique means for studying galaxy number counts at the fainter end \citep{1997ApJ...490L...5S, 2002AJ....123.2197C, 2008MNRAS.384.1611K, 2011A&A...527A.117J,2013ApJ...776..131C,2016ApJ...829...25H, 2022ApJ...939....5C}.
Gravitational lensing occurs when the gravitational field of a massive object, such as a rich cluster of galaxies, bends and magnifies the light from distant galaxies. By exploiting the lensing effect, we can effectively boost the observed flux of background galaxies, allowing us to probe deeper into the Universe and uncover fainter sources that would otherwise remain undetected in blank-field surveys.

In this study, we utilize the deep data obtained with SCUBA-2 on 10 massive galaxy cluster fields to derive robust number counts at 450\,$\micron$. Our analysis properly accounts for selection biases, completeness, and uncertainties. By constructing the deepest 450\,$\mu$m number counts ever, we aim to unravel the relative contributions of SMGs to the 450\,$\mu$m EBL. In Section \ref{sec:data}, we describe our SCUBA-2 data and data reduction. In Section~\ref{sec:methodology}, we describe our methodology, including source extraction, Monte Carlo simulations, and number counts calculations. In Section \ref{sec:Discussion}, we present our number counts and the integrated energy density. We summarize our results in Section~\ref{sec:Summary}.

% \section{Data} \label{sec:data}
\section{Data and Data Reduction} \label{sec:data}
We retrieved the SCUBA-2 450\,$\mu$m data from the CADC archive. We used the data taken under weather band 1 and band 2 conditions ($\tau_{225~\rm GHz}~<$ 0.08) between October 2011 and July 2022 \citep{2022ApJ...939....5C}. The scan pattern used for these 10 cluster fields was CV DAISY, which has a roughly circular field size of $\simeq6^{\prime}$ in radius. We summarize the data in Table \ref{table:scuba2data}.

We reduced the data following \citet{2013ApJ...776..131C}. We used the Dynamic Iterative Map Maker (DIMM) method in the SMURF package contained in the STARLINK software \citep{2013MNRAS.430.2545C}. This method models individual components that make up the time-series recorded by the bolometer to produce science maps. We adopted the ``blank field" configuration file, which is suitable for detecting faint point sources in extragalactic surveys. For each cluster field, we produced scan maps with a pixel scale of $1''$ and then applied the recommended Flux Conversion Factors (FCFs) from \citet{2021AJ....162..191M} to convert the pixel unit from picowatts to Jansky per beam. After calibration, we used the \texttt{MOSAIC\_JCMT\_IMAGES} recipe from the Pipeline for Combing and Analyzing Reduced Data (PICARD) to co-add and mosaic calibrated scans for each field. 

To improve source detection, we applied a matched filter using the \texttt{SCUBA2\_MATCHED\_FILTER} recipe in PICARD. This recipe first convolves the map with a Gaussian to estimate the low spatial frequency noise and then subtracts it from the original map. We adopted the default $20^{\prime \prime}$ FWHM value for the Gaussian profile. We verified the flux recovery capability of \texttt{SCUBA2\_MATCHED\_FILTER} following \citet{2020ApJ...889...80L}, and we adopted a mean upward correction of $5.3\%$ for the flux loss to our 450\,$\mu$m data.

\begin{deluxetable}{ccccc}
\centering
% This sets the table to its 'natural' width, i.e. not trying to
% expand to the width of the text:
\tablewidth{0pt}  
\tablecaption{SUMMARY of SCUBA-2 450 $\micron$ DATA\label{table:scuba2data}
}
\tablehead{
\colhead{Field} & \colhead{RA} &
\colhead{Dec} & \colhead{Exposure}&\colhead{Central RMS}\\
\colhead{} &\colhead{}&\colhead{}&\colhead{(hours)} & \colhead{(mJy)} }
\startdata
A370  & 39.9604   & -1.5856   & 29.1 & 2.19 \\
A1689 & 197.8729  & -1.3411   & 27.7 & 2.16 \\
A2390 & 328.3979  & 17.6867   & 39.6 & 2.28 \\
A2744 & 3.5788    & -30.3894  & 24.5 & 3.08 \\
MACS~J0416.1-2403 & 64.0349   & -24.0724  & 22.1 & 2.42 \\
MACS~J0717.5+3745 & 109.4020  & 37.7564   & 51.5 & 1.51 \\
MACS~J1149.5+2223 & 117.3962  & 22.4030   & 35.0 & 1.46 \\
MACS~J1423.8+2404 & 215.9486  & 24.0778   & 43.8 & 1.76 \\
MACS~J2129.40741  & 322.3592  & -7.6906   & 12.6 & 4.07 \\
RX~J1347.5-1145   & 206.8775  & -11.7528  & 20.0 & 2.27 \\
\enddata
\end{deluxetable}

\begin{figure*}[t]
    \centering

    \begin{minipage}{\columnwidth}
    \includegraphics[width=1.0\linewidth]{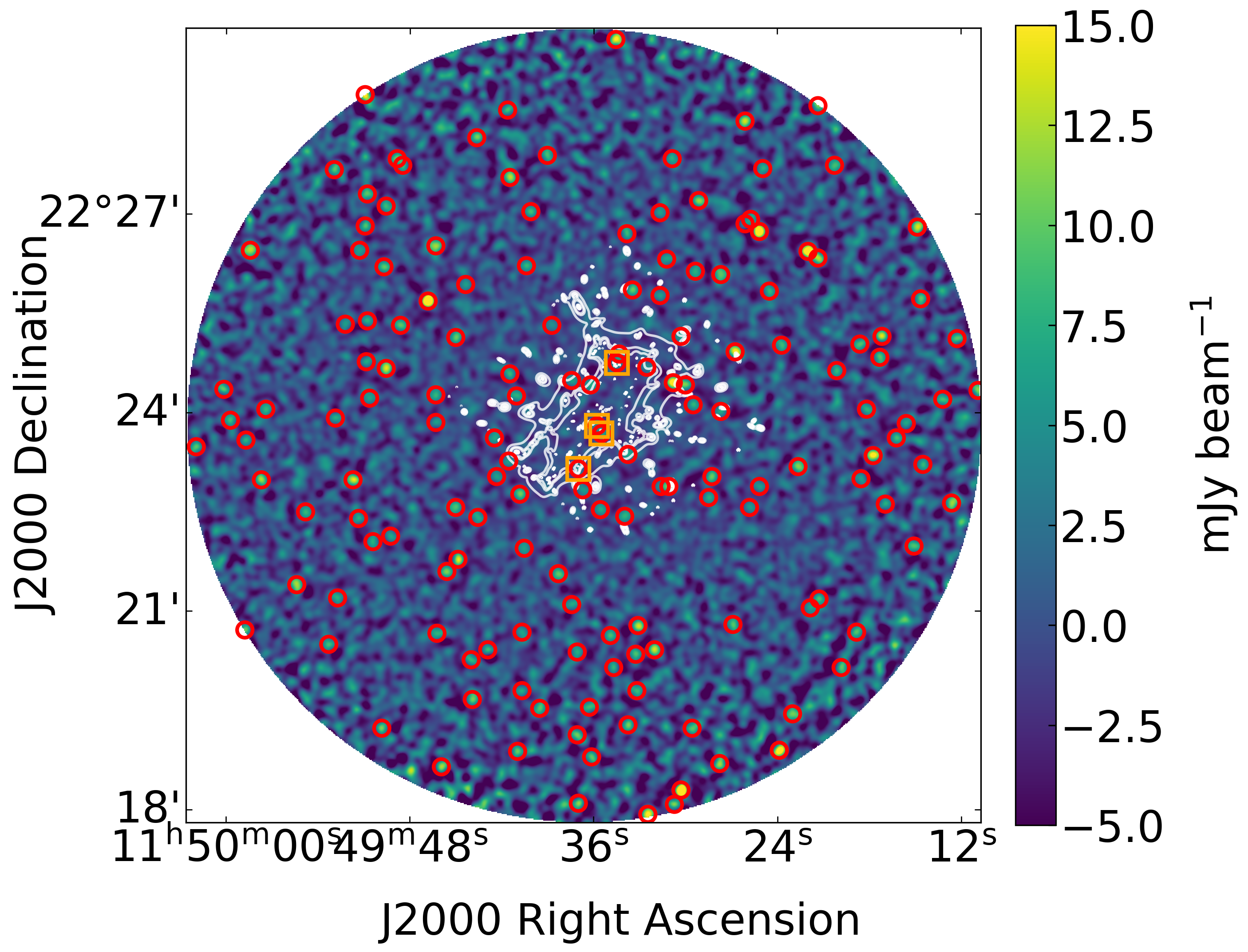}
    \end{minipage}\hfill
    \begin{minipage}{\columnwidth}
    \includegraphics[width=1.0\linewidth]{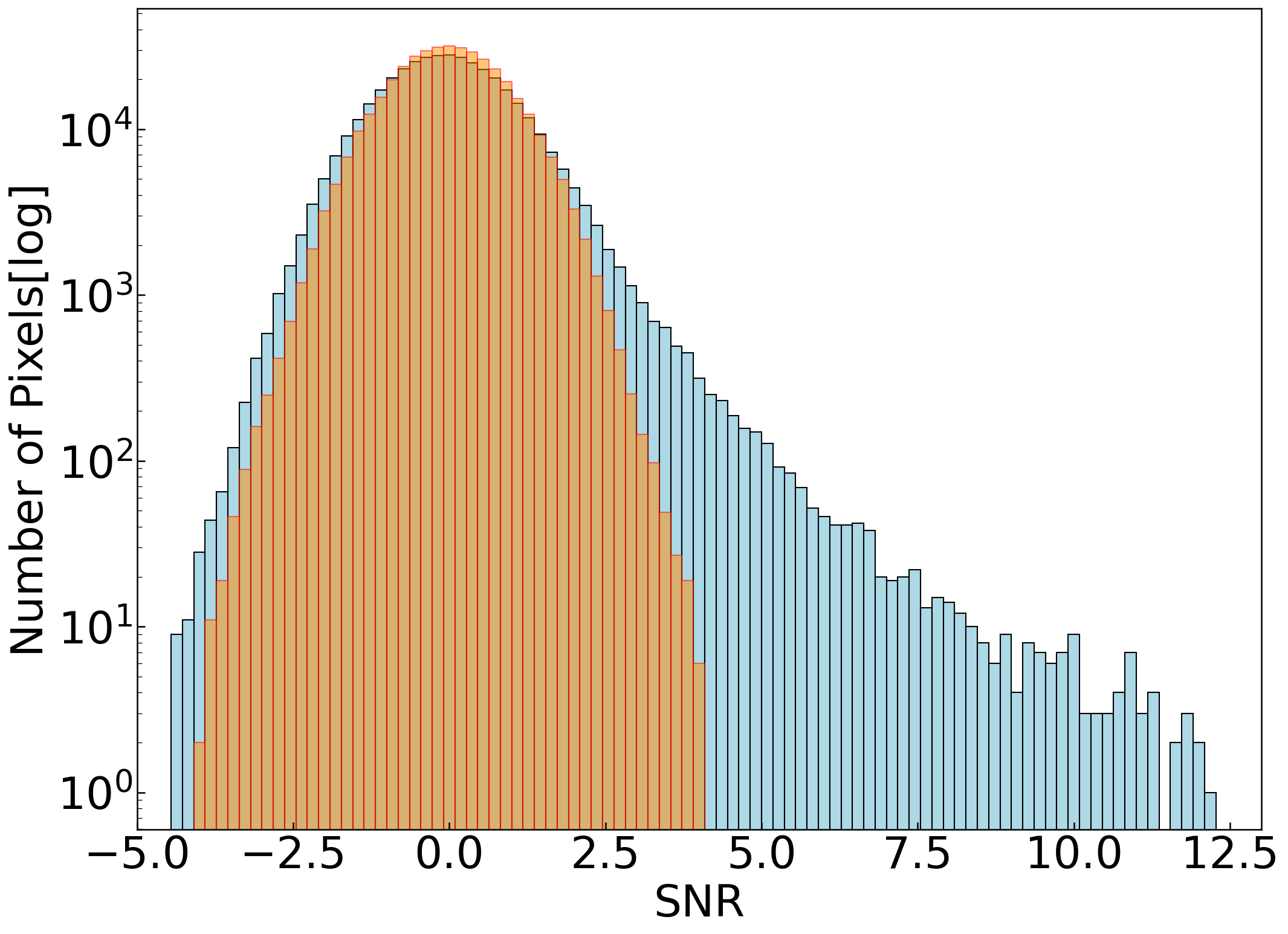}
    \end{minipage}

    \caption{{\it Left:} An example SCUBA-2 450\,$\micron$ flux density map of the cluster MACS J1149.5+2223 with a 6$^\prime$ radius circular footprint. The red circles on the 450\,$\mu$m map show the extracted $>3\sigma$ sources. The orange boxes mark the sources below 1\,mJy after delensing. The white contours represent the CATS lens model for $z = 1.5$ at magnification values of 1.5, 2.0, and 2.5 (moving inwards). {\it Right:} Histogram of the SNR values based on the map of MACS J1149.5+2223. The orange region shows the SNR distribution in the jackknife map. The blue represents detections in the data map.\label{fig:source extraction 1149}}
\end{figure*}

% \section{Methodology} \label{sec:methodology}
\section{Analysis} \label{sec:methodology}

\subsection{Source Extraction}
Before source extraction, for each field we generated SCUBA-2 PSF models by stacking the 10-20 highest signal-to-noise ratio (SNR) source images without neighboring sources. This method inherently assumes that the typical source size is much smaller than the beam size of SCUBA-2 in the submillimeter ($\sim7\farcs5$ at 450\,$\micron$), which is supported by recent ALMA observations that found typical sizes of subarcseconds (e.g., \citealt{2015ApJ...807..128S, 2016ApJ...833..103H, 2017ApJ...850...83F, 2018ApJ...859...12G, 2020ApJ...901...74T}). 
We then fitted a double Gaussian profile to model the PSFs, and we used the best-fit model PSFs for source extraction. 

We performed the source extractions as in \citet{2016ApJ...829...25H}. We searched for the maximum SNR pixel in the central circular region ($6^{\prime}$ radius). (The pointing centers of the maps are normally close to the cluster centroids.) We recorded the location and flux density of the pixel, then subtracted a rescaled PSF centered at this pixel and searched for the next maximum SNR pixel.
Following \citet{2016ApJ...829...25H}, we used a 3$\sigma$ threshold, which allows us to obtain a better SNR in the number counts at the faint end. We repeated the extraction process until we reached this threshold. In Figure \ref{fig:source extraction 1149}, we show one of our cluster fields as an example for the source extraction.

\subsection{Delensed Raw Number Counts}
We calculated delensed differential number counts at 450\,$\micron$ by using sources with SNR $> 3.0$. To compute the demagnified flux densities, we used the public software LENSTOOL \citep{2011ascl.soft02004K} to generate magnification maps for our cluster fields. Since we do not have redshifts for these sources, we corrected the flux densities of the sources by adopting estimated median redshifts of 1.5 based on previous 450\,$\micron$ studies (e.g., \citealt{2013ApJ...776..131C, 2013MNRAS.436.1919C}). We took our lens models from the LENSTOOL developers and Hubble Frontier Fields Archive (\citealt{2014ApJ...781....2A, 2009ApJ...706.1201B, 2017A&A...600A..90C, 2015MNRAS.446..683D, 2015ApJ...800...38G, 2016ApJ...831..182H, 2015ApJ...799...12I, 2007NJPh....9..447J, 2009MNRAS.395.1319J, 2012MNRAS.426.3369J, 2014MNRAS.443.1549J, 2014ApJ...797...48J, 2010GReGr..42.2151K, 2018ApJ...855....4K, 2006MNRAS.367.1209L, 2017ApJ...837...97L, 2011MNRAS.417..333M, 2014MNRAS.443.3631M, 2014MNRAS.439.2651M, 2010PASJ...62.1017O, 2014MNRAS.444..268R, 2013ApJ...762L..30Z}).

We determined the magnification factors $\mu_{\rm i}$ of the point sources from the magnification maps. The demagnified flux density of each source can be obtained from 
\begin{equation}
    S_{\rm demag, \emph i} = \frac{S_{\rm obs, \emph i}}{\mu_{i}} \,,
\end{equation}
 where $S_{\rm demag, \emph i}$ and $S_{\rm obs, \emph i}$ are demagnified and observed flux densities, respectively.
We calculated the effective area A$_{\rm eff,\emph i}$ of each source on the source plane. We summed over the pixels whose SNRs were greater than 3$\sigma$ and then converted pixels to square degrees. We then calculated the delensed raw counts at the $j$–th flux bin as
\begin{equation}
    \frac{dN_{\rm raw, j}}{dS_j} = \frac{1}{\Delta S_j}\sum_{i}^{n} X_{i}
\end{equation}
and
\begin{equation}
    X_{i} = \frac{1}{A_{\rm eff, \emph i}} \,.
\end{equation}
Here, X$_{i}$ represents the number density contribution of each source within that flux bin. We based the error calculation on Poisson statistics. 

\subsection{Simulations}

Corrections for flux boosting, false detections, and incompleteness are needed in order to obtain the intrinsic number counts. 
To do this, we ran Monte Carlo simulations to find the underlying models for our fields. We used the Schechter function form as our number counts model:
\begin{equation}\label{eq:4}
    \rm \frac{dN}{dS} = \Bigl(\frac{N_{0}}{S_{0}}\Bigr)~\Bigl(\frac{S}{S_{0}}\Bigr)^{\alpha}~exp\Bigl(-\frac{S}{S_{0}}\Bigr) \,.
\end{equation}

We generated artificial sources whose flux densities were assigned according to the underlying models for each cluster field. We then randomly distributed these sources in the source plane. Next, we projected the simulated sources onto the image plane using LENSTOOL. The outputs of LENSTOOL contain the new flux densities and positions of the simulated sources in the image plane. We convolved the simulated sources with the PSF and added them into the jackknife maps to produce mock observation maps. We produced jackknife maps following \citet{2016ApJ...829...25H} by coadding even and odd scans separately. We then subtracted these two coadded maps and rescaled the value of each pixel by a factor of $\sqrt{t_{\rm even}} \times \sqrt{t_{\rm odd}}$/(t$_{\rm even}$ + t$_{\rm odd}$), where t$_{\rm even}$ and t$_{\rm odd}$ represent the integration times of each pixel from the even and odd coadded maps, respectively. We also applied matched-filtering to the jackknife maps, as we did for the real data images.

In order to estimate the true number counts, we adopted an iterative procedure in our simulations. We generated 15–20 mock maps in each iteration step. We then performed source extraction and computed the averaged recovered counts of the mock maps. Next, we corrected the raw counts by using the ratio between the averaged recovered counts and the raw counts. Finally, we did a $\chi^{2}$ fit to the corrected counts by using the Schechter function. This fit will be the new counts model for the next iteration. We iterated until the recovered counts converged with the raw counts to within the 1$\sigma$ uncertainties for all the flux bins.

Once we obtained the intrinsic number counts models, we used these to produce 500 mock images and then performed the source extractions to create the source catalogs. Following \citet{Gao2024aa}, we cross-matched the input and output catalogs within a 1/2 beam FWHM as our search area to find the brightest counterparts. We considered S$_{\rm output}$/S$_{\rm input}\leq3$ a match in this analysis, which is similar to what has been adopted previously (e.g., \citealt{2017MNRAS.465.1789G}).
Following \citet{Gao2024aa}, we estimated the boosting factors, false detection rates, and completeness of the point sources using a two–dimensional binning method in our cross-matched catalogs.

\subsection{Delensed Corrected Number Counts}
To derive delensed corrected number counts, we first deboosted the fluxes of the sources. We then delensed their fluxes and estimated their effective areas in the source plane. 
Finally, we derived the delensed corrected number counts using%
\begin{equation}
    \frac{dN_{\rm corr, j}}{dS_j} = \frac{1}{\Delta S_j}\sum_{i}^{n} X_{\rm corr,\emph i}
\end{equation}
and
\begin{equation}
    X_{\rm corr, \emph i} = \frac{1-p_{\rm false, \emph i }}{C_{i}A_{\rm eff,\emph i}} \,,
\end{equation}
where p$_{\rm false, \emph i }$ is the false detection rate, and C$_{i}$ is the completeness. We confirmed that the delensed corrected number counts agree with the intrinsic counts models.
We estimated the statistical uncertainties from the source positions and took these uncertainties into account in the analyses.

% \section{Results} \label{sec:result}
\subsection{Systematic Uncertainties} \label{sec:result}
Systematic uncertainties related to the modeling of gravitational lensing need to be taken into account for a proper assessment of the number counts error budget. In the following, we give our results with estimates of the systematic uncertainties caused by the redshift distribution of the background lensed SMGs, the lens models, and the clustering of the source distributions.

In our methodology, we calculated corrected number counts by assuming a median redshift of 1.5. However, it is expected that the SMGs have a redshift distribution, which could affect the magnification estimates. This leads to additional uncertainties in the number counts.
To address this, we randomly assigned redshifts to our sources using the redshift distribution from the STUDIES survey, which is the deepest 450\,$\micron$ blank-field survey (\citealt{2017ApJ...850...37W, 2021MNRAS.500..942D, Gao2024aa}). We calculated the corrected number counts again and compared the standard deviations of the number counts in the randomized redshift sample with the Poisson noise.
We found that the uncertainties caused by the assumptions concerning the redshift are subdominant, and they are, on average, about 25\% of the Poisson uncertainties. Nevertheless, we include this error budget in the total error budget.

\begin{deluxetable}{cccc}
\tablewidth{0pt}  
\tablecaption{Combined Differential Number Counts at 450\,$\micron$\label{table:combinedcounts}
}
\tablehead{
\colhead{S$_{450~\micron}$} & \colhead{log$_{10}$(\rm dN/dS)}& \colhead{$<\mu>^1$}& \colhead{N$_{\rm total}^2$}\\
\colhead{(mJy)} &\colhead{(mJy$^{-1}$~deg$^{-2}$)} &\colhead{} &\colhead{}}
\startdata
0.13 & 6.2$_{-1.4}^{+0.4}$ & 34.4 & 3\\
0.25 & 5.6$_{-99.9}^{+0.6}$ & 22.4 & 5\\
0.45 & 5.1$_{-0.7}^{+0.3}$ & 14.9 & 12\\
0.81 & 4.7$_{-99.9}^{+0.5}$ & 7.4 & 3\\
1.48 & 4.2$_{-0.5}^{+0.3}$ & 4.1 & 26\\
2.70 & 3.5$_{-0.2}^{+0.2}$ & 2.1 & 108\\
4.91 & 3.0$_{-0.1}^{+0.1}$ & 1.6 & 232\\
8.94 & 2.4$_{-0.1}^{+0.1}$ & 1.2 & 358\\
16.28 & 1.5$_{-0.2}^{+0.2}$ & 1.1 & 228\\
29.65 & 0.6$_{-0.8}^{+0.4}$ & 1.2 & 23\\
\enddata
\tablenotetext{1}{Averaged gravitational magnification}
\tablenotetext{2}{Total number of sources in each flux bin}
\end{deluxetable}

To estimate the systematic uncertainties caused by the lens modeling, we utilized the various lens models provided for the five Hubble Frontier Fields (A370, A2744, MACS J0416.1-2403, MACS J0717.5+3745, and MACS J1149.5+2223). We ran the same Monte Carlo simulations as those done on the real data but using different lens models. We then estimated the systematic uncertainties by calculating the standard deviations of the corrected counts obtained using the different lens models. We found the systematic uncertainties to be subdominant, again about 25\% of the Poisson uncertainties. We include this error budget in the total error budget.

We note that in our Monte Carlo simulations, we randomly distributed the positions of the artificial sources. This might bias our counts, since we do not take clustering effects into account when we calculate the boosting factors, false detection rates, and completeness from the mock maps. To test whether neglecting clustering could significantly alter our counts results, we used the empirical catalogs produced by the SIDES simulation \citep{2017A&A...607A..89B}, which inherently include clustering effects, since the simulation builds upon dark matter lightcones.

\begin{figure}[ht!]
\includegraphics[width=0.45\textwidth]{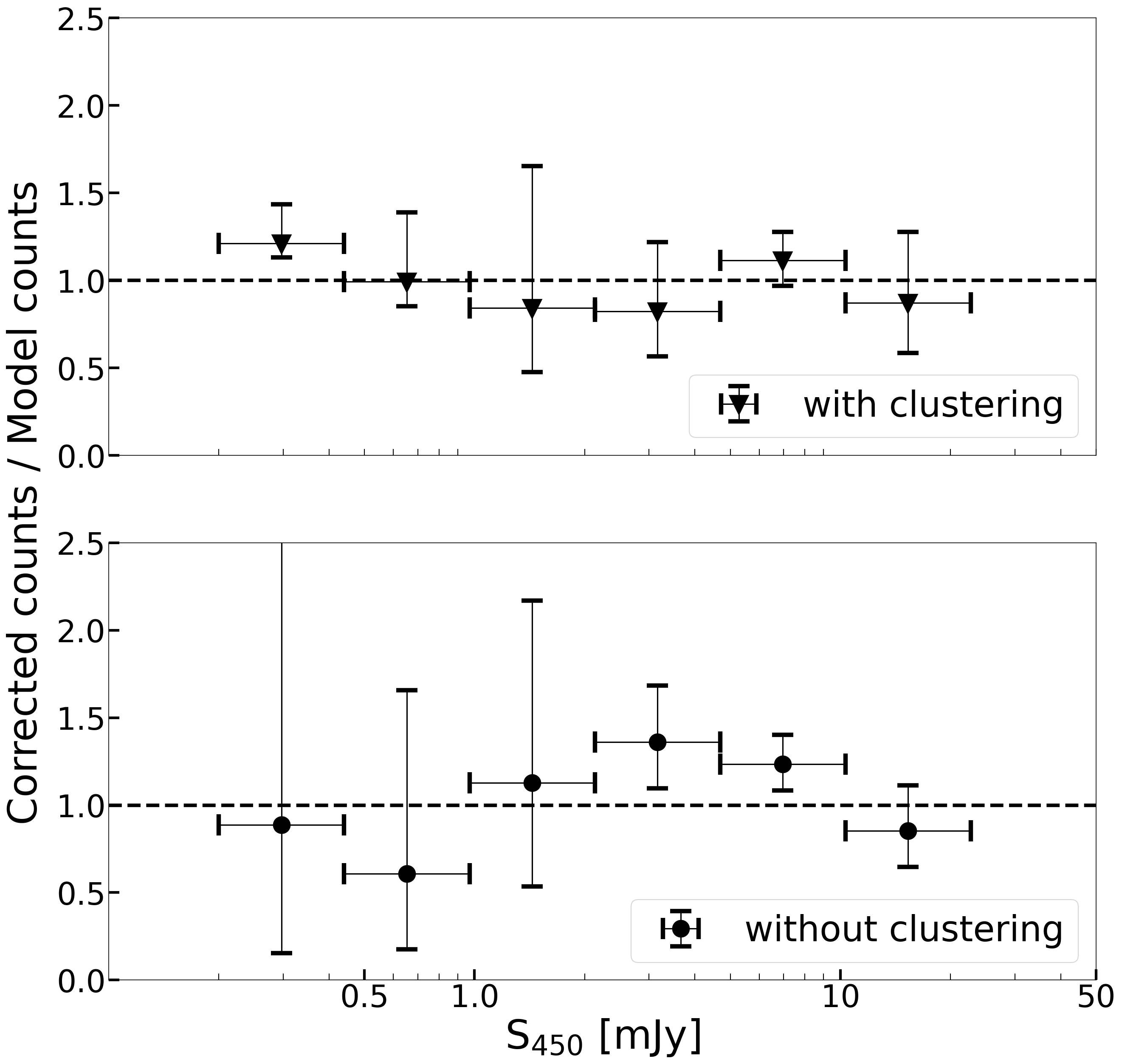}
\caption{Ratios between the corrected number counts and the model counts, where the model counts are based on the SIDES simulation source catalog, and the corrected counts were obtained by applying our methodology to the mock images that were made based on the SIDES source catalog. The top panel shows the results for the case where the source positions are adopted from SIDES, which includes clustering. The bottom panel shows the results for the case where the sky positions of the sources are randomized.
}
\label{fig:SIDES corr counts}
\end{figure}

Before assessing the clustering effects, we first validated our methodology for estimating intrinsic counts by performing the following test. We clipped the original SIDES 2~deg$^2$ simulated map into a set of 25 smaller cutout maps with a size similar to our SCUBA-2 footprint, and we treated them as different cluster fields in the source planes. We then lensed these simulated maps to the image planes using LENSTOOL and used our methodology to find the corrected counts. We compared the corrected counts to the true counts provided by SIDES. In this test, we adopted the three-dimensional source positions from the SIDES catalog instead of having the sources randomly distributed on the sky with an assigned redshift. We show our results in Figure~\ref{fig:SIDES corr counts}, where the averaged corrected number counts over the 25 cutout maps are consistent with the true number counts. 

After validating our methodology, we moved on with a similar test. This time we randomly distributed the source positions to calculate the corrected number counts using the smaller SIDES cutout maps. We show our results in Figure \ref{fig:SIDES corr counts}. While not taking clustering into account can lead to larger uncertainties, evidently with a larger scatter in the averaged counts, on average, there is no significant difference between the SIDES counts and the corrected counts. Our results are therefore consistent with previous studies---either from other SCUBA-2 450\,$\micron$ surveys \citep{2017ApJ...850...37W}, or from Herschel studies with similar beam sizes but at slightly shorter wavelengths \citep{2015A&A...573A.113B}---where no significant impact of clustering on the number counts was found.

\section{Results and Discussion} \label{sec:Discussion}
\begin{figure*}[t]
    \centering

    \begin{minipage}{\columnwidth}
    \includegraphics[width=1.0\linewidth,height=0.8\linewidth]{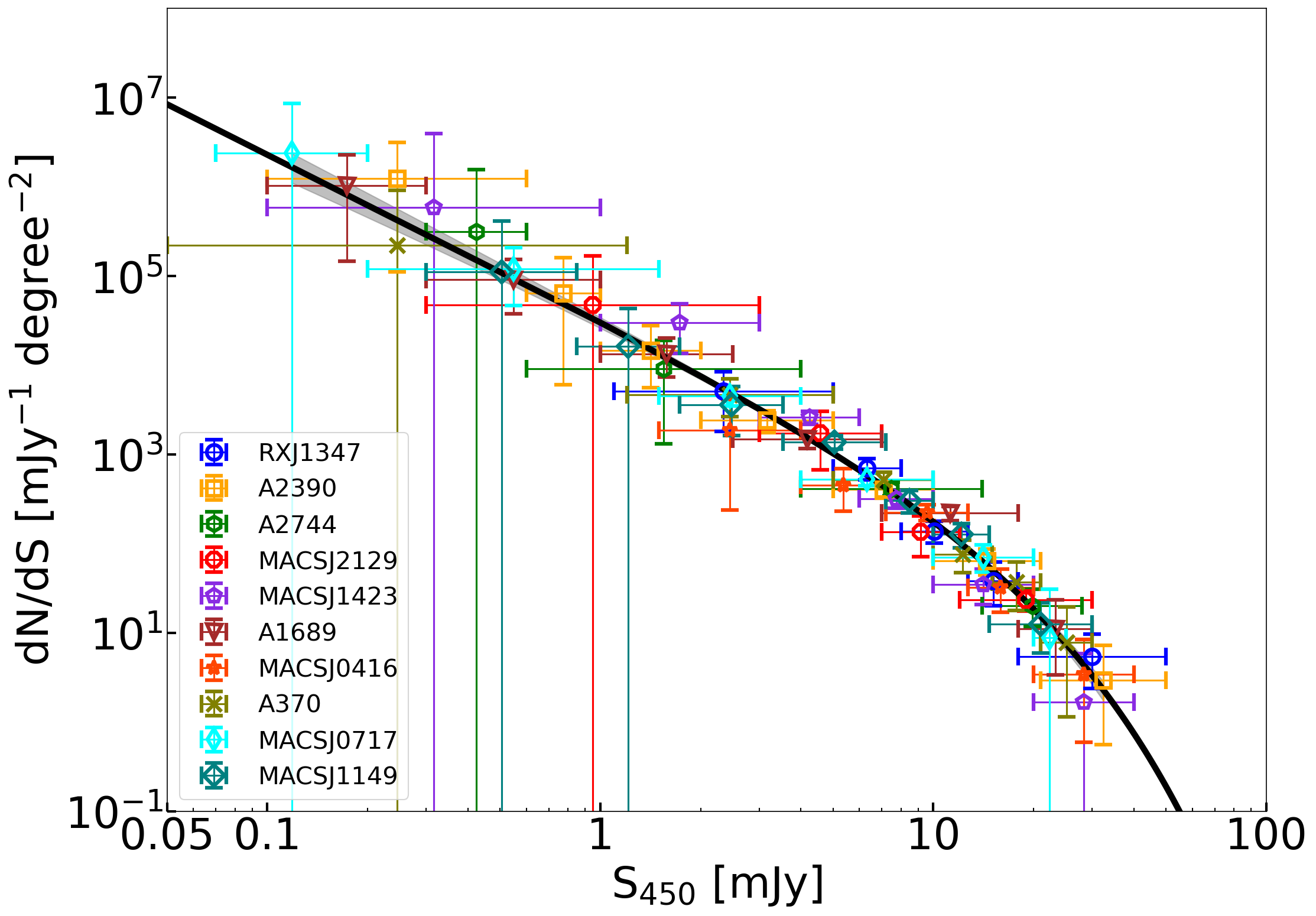}
    \end{minipage}\hfill
    \begin{minipage}{\columnwidth}
    \includegraphics[width=1.0\linewidth,height=0.8\linewidth]{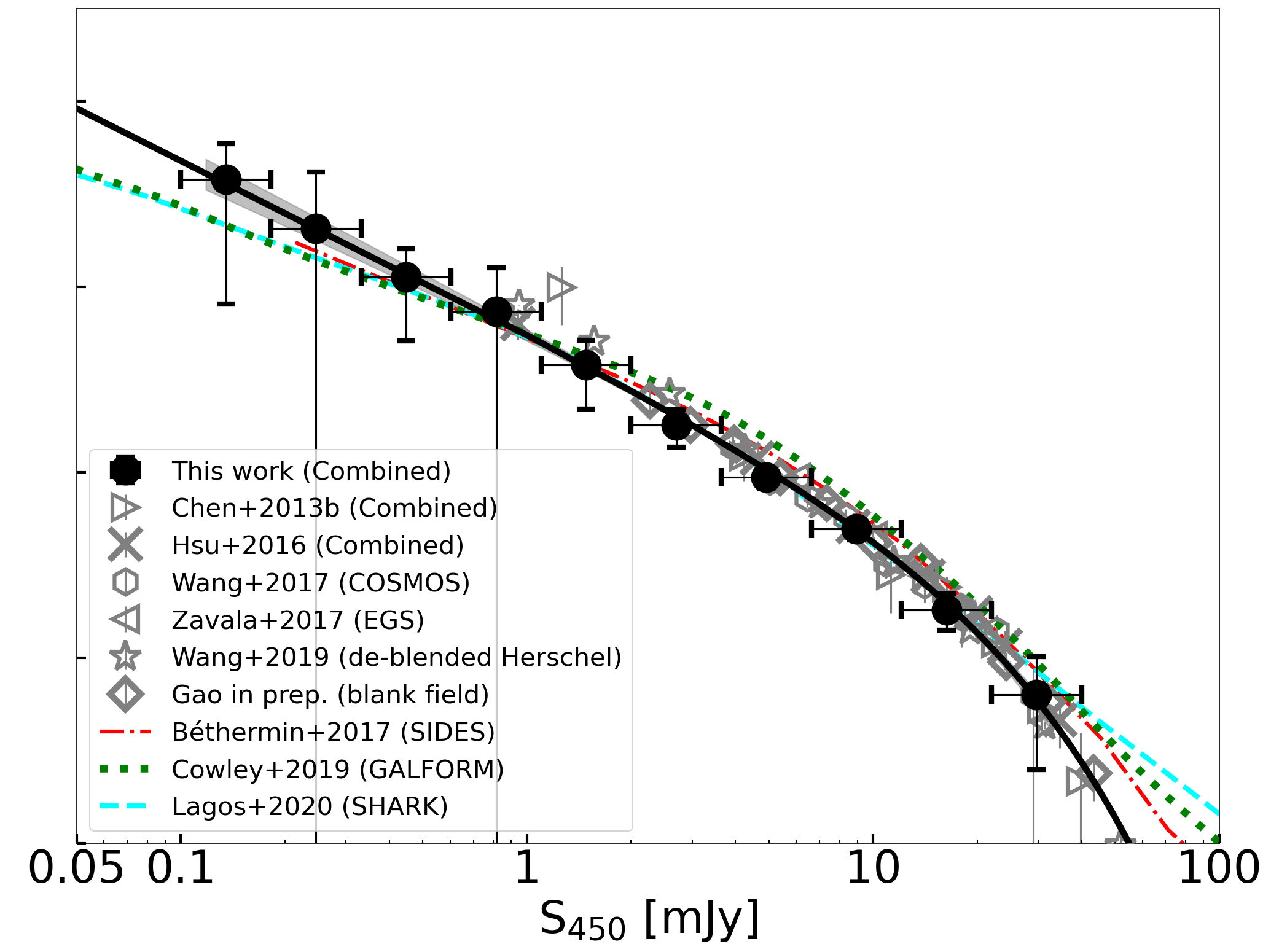}
    \end{minipage}
\caption{
{\it Left:} Corrected differential number counts for all 10 cluster fields at 450\,$\mu$m. The solid black curve shows the best-fit Schechter function, while the shaded gray region is the 68\% confidence interval. The uncertainties include Poisson noise, as well as those caused by the underlying redshift distributions and lens models. {\it Right:} Weighted average counts of all 10 cluster fields in filled black circles, along with previous 450\,$\micron$ surveys in open symbols (\citealt{2013ApJ...776..131C, 2016ApJ...829...25H, 2017MNRAS.464.3369Z, 2017ApJ...850...37W, Gao2024aa}). Model predictions from SIDES \citep{2017A&A...607A..89B}, GALFORM \citep{2019MNRAS.487.3082C}, and SHARK \citep{2020MNRAS.499.1948L} are shown as dotted and dashed curves. 
\label{fig:delensed corr counts}}
\end{figure*}

We present our corrected differential number counts for the 10 lensing cluster fields in the left panel of Figure~\ref{fig:delensed corr counts}. The solid black curve is the best-fit Schechter function for our corrected counts, which can be parameterized by Equation~\ref{eq:4} with the following parameters: N$_{0}$ = 4437.5$\pm$1399.5, S$_{0}$ = 10.4$\pm$1.7, and $\alpha$=-1.9$\pm$0.1. The uncertainties on each data point include Poisson noise and the uncertainties from the redshift distribution and the lens model.
Thanks to the powerful effects of strong gravitational lensing, we have pushed the detection limit down to $\sim$0.1\,mJy at 450\,$\micron$, a factor of $>$10 improvement over the deepest blank-field counts (\citealt{2017ApJ...850...37W, Gao2024aa}).

We also calculate the weighted average counts by combining all 10 fields. We show the results in the right panel of Figure~\ref{fig:delensed corr counts}, and we provide the corresponding values in \autoref{table:combinedcounts}.
Our results are consistent with the previous measurements shown in Figure~\ref{fig:delensed corr counts}. However, the various physical or empirical models (dotted and dashed curves in the right panel) tend to overpredict source densities at the bright end at 450\,$\micron$ ($\gtrsim$1\,mJy) by about 10-30\%. On the other hand, at the faint end ($\lesssim$1\,mJy), our counts are slightly higher, although not significantly. The physical reasons for the disagreement between the measurements and the models at the bright end need to be investigated. \citet{Gao2024aa} proposed that it could be due to the mismatch in halo masses between those inferred from clustering measurements \citep{2020ApJ...889...80L} and those in the models. Studies of the physical properties of the 450\,$\micron$ sources, such as stellar mass, could potentially shed more light on this issue. 

\begin{figure}[ht!]
\includegraphics[width=0.45\textwidth]{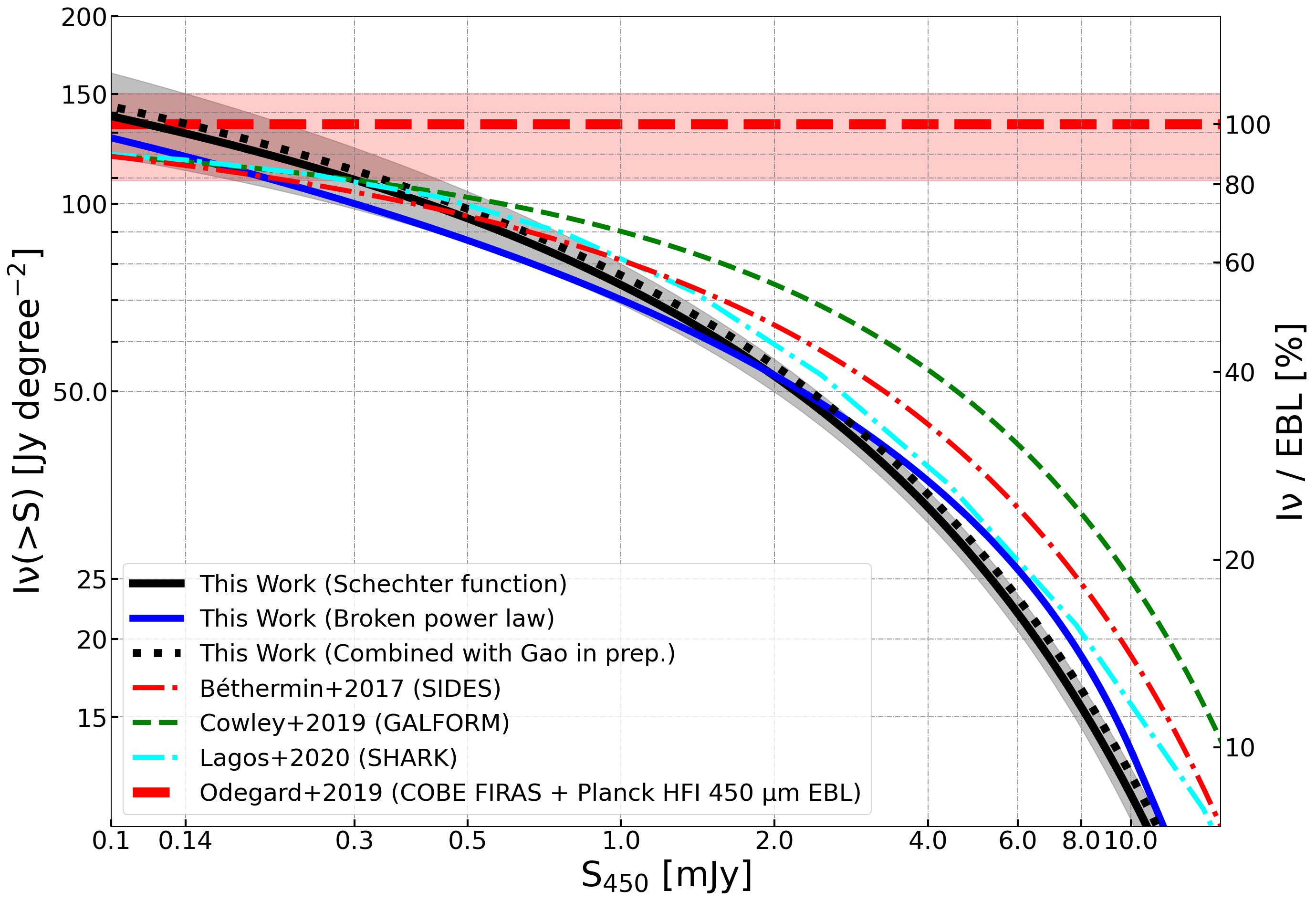}
\caption{Cumulative EBL as a function of flux density at 450\,$\mu$m. The black solid curve was calculated based on our best-fit Schechter function with the 1$\sigma$ uncertainties in gray shading. The blue solid curve was calculated based on a broken power law function. The black dotted curve was calculated by combining our results with the deepest blank-field counts from \citet{Gao2024aa}.
The horizontal red dashed line is the EBL measured by the COBE$+$Planck satellites
\citep{2019ApJ...877...40O}. The red shaded region shows the range from COBE estimates (e.g., \citealt{1996A&A...308L...5P, 1998ApJ...508..123F, 2000A&A...360....1G}).\label{fig:EBL} }
\end{figure}

To estimate the contributions of the detected sources to the 450\,$\micron$ EBL, we integrated the best-fit Schechter model, which we show as a function of flux in Figure~\ref{fig:EBL}. Above 0.1\,mJy where significant counts are obtained, we find the total energy density at 450\,$\mu$m to be 138.1$^{+23.9}_{-19.3}$~Jy~deg$^{-2}$, or 0.45$^{+0.08}_{-0.06}$~MJy~sr$^{-1}$. This corresponds to 103$^{+18}_{-14}\%$ of the total 450\,$\mu$m EBL reported by \citet{2019ApJ...877...40O}. 

Past works have suggested that a broken power law could be a viable alternative to the Schechter function for the underlying counts models. It can be described as \\
\begin{equation}
\frac{dN}{dS}=
    \begin{cases}
      N_{0}(\frac{S}{S_{0}})^{-\alpha}, & if\ S\leq S_{0} \\
      N_{0}(\frac{S}{S_{0}})^{-\beta}, & if\ S > S_{0}
    \end{cases}
\end{equation}
\\
We ran further counts analyses based on this model form and found consistent results compared to those obtained based on Schechter functions. The best-fit parameters of the broken power law are N$_{0}$=151.0$\pm$14.6, S$_{0}$=9.6$\pm$0.4, $\alpha$=2.0$\pm$0.1, $\beta$=5.6$\pm$1.1. We plot the cumulative energy density of the best-fit broken power law in Figure 4. The cumulative energy density based on the broken power law is 126.9$^{+41.6}_{-41.6}$ Jy~deg$^2$.

Our work demonstrates for the first time that discrete sources, are the dominant contributors to the 450\,$\micron$ EBL. Interestingly, about half of the contribution comes from sources that are fainter than $\sim$1\,mJy, below the typical confusion limit of SCUBA-2 450\,$\micron$ images \citep{Gao2024aa}.

Noticeably, as shown in \autoref{fig:EBL}, the integrated energy density does not converge when integrating down to 0.14\,mJy, suggesting that deeper data are needed in order to put a tighter constraint on the faint end counts and thus obtain a converged constraint on the EBL contributions from discrete sources.

\section{Summary} \label{sec:Summary}
In summary, our research represents a significant step forward in the study of the 450\,$\micron$ EBL. Through the innovative combination of gravitational lensing and the unparalled capability of SCUBA-2, we have achieved a complete resolution of the 450\,$\micron$ EBL and established the dominance of discrete SMGs as its main contributors. These findings provide a broader understanding of the FIR/submillimeter regime, the cosmic energy distribution, and the interplay between galaxies and the diffuse background radiation. Our measurements could also be helpful for the design of the next generation submillimeter facilities, such as LST \citep{Kohno_2020} and AtLAST \citep{2020SPIE11445E..2FK}, which aim to obtain wide blank-field images to a depth that is similar to what has been reached by this work.

\begin{acknowledgments}
We thank the reviewer for a useful report that has improved the manuscript. Q.-N.H. and C.-C.C acknowledge support from the National Science and Technology Council of Taiwan (NSTC 109-2112-M-001-016-MY3 and 111-2112M-001-045-MY3), as well as Academia Sinica through the Career Development Award (AS-CDA-112-M02). L.L.C acknowledges support from NASA grant 80NSSC22K0483.
A.J.B. acknowledges support from a Kellett Mid-Career Award and a WARF Named Professorship from the University of Wisconsin-Madison Office of the Vice Chancellor for Research and Graduate Education with funding from the Wisconsin Alumni Research Foundation. The James Clerk Maxwell Telescope is operated by the East Asian Observatory on behalf of The National Astronomical Observatory of Japan; Academia Sinica Institute of Astronomy and Astrophysics; the Korea Astronomy and Space Science Institute; the National Astronomical Research Institute of Thailand; Center for Astronomical Mega-Science (as well as the National Key R\&D Program of China with No. 2017YFA0402700). Additional funding support is provided by the Science and Technology Facilities Council of the United Kingdom and participating universities and organizations in the United Kingdom and Canada. Additional funds for the construction of SCUBA-2 were provided by the Canada Foundation for Innovation. The authors wish to recognize and acknowledge the very significant cultural role and reverence that the summit of Maunakea has always had within the indigenous Hawaiian community.  We are most fortunate to have the opportunity to conduct observations from this mountain.
This work utilizes gravitational lensing models produced by PIs Bradač, Natarajan \& Kneib (CATS), Merten \& Zitrin, Sharon, Williams, Keeton, Bernstein and Diego, and the GLAFIC group. This lens modeling was partially funded by the HST Frontier Fields program conducted by STScI. STScI is operated by the Association of Universities for Research in Astronomy, Inc. under NASA contract NAS 5-26555. The lens models were obtained from the Mikulski Archive for Space Telescopes (MAST).
\end{acknowledgments}

\vspace{5mm}
\facilities{JCMT(SCUBA-2)}

\software{astropy (\citealt{2013A&A...558A..33A, 2018AJ....156..123A}), LENSTOOL \citep{2011ascl.soft02004K}}

\appendix

\section{Appendix A}
\begin{figure}[ht!]
\centering
\includegraphics[width=1.0\textwidth]{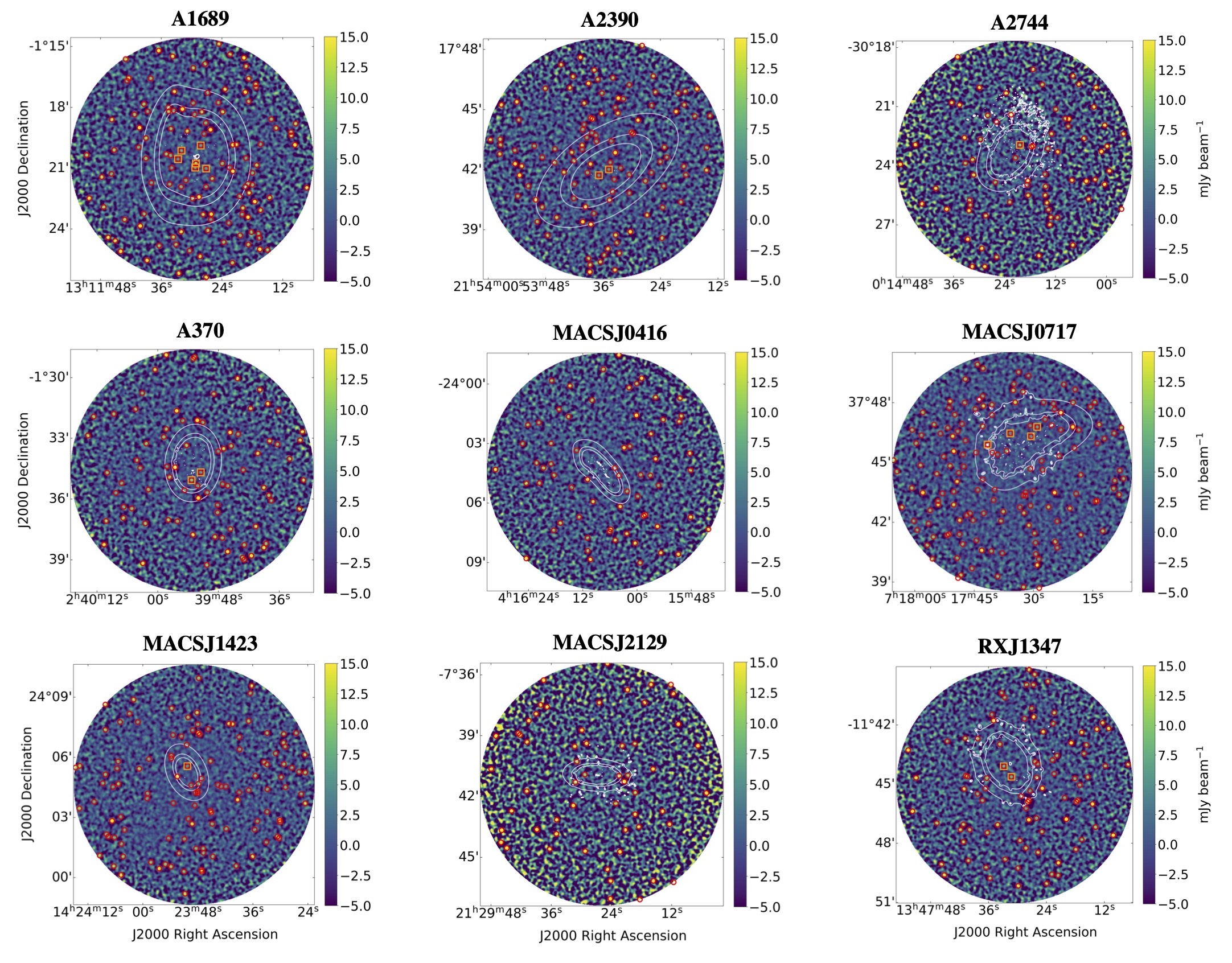}
\caption{Flux density maps for the remaining nine fields. The format follows that adopted in the left panel of Figure \ref{fig:source extraction 1149}.\label{fig:overview}}
\end{figure}

In \autoref{fig:overview} we show the imaging data for the remaining nine fields. The symbols follow those used in Figure \ref{fig:source extraction 1149}.

\bibliography{draft}
\bibliographystyle{aasjournal}

\end{document}

\begin{figure}
     \centering
     \begin{subfigure}[ht]{0.32\textwidth}
         \centering
         \textbf{A1689}\par\medskip
         \vspace{-2mm} 
         \includegraphics[width=\textwidth]{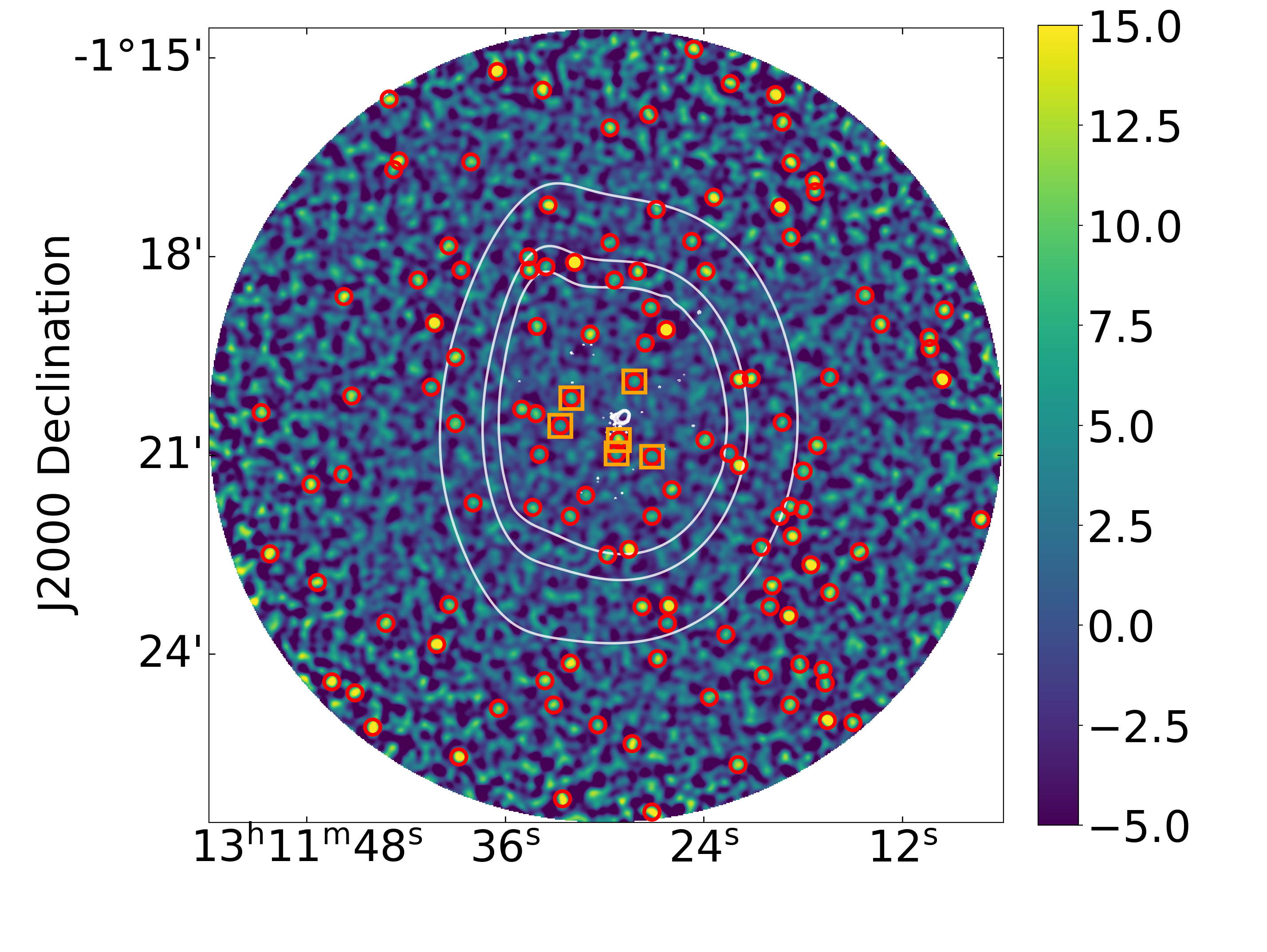}

         \label{fig:A1689}
     \end{subfigure}
     \hfill
     \begin{subfigure}[ht]{0.32\textwidth}
         \centering
         \textbf{A2390}\par\medskip
         \vspace{-2mm} 
         \includegraphics[width=\textwidth]{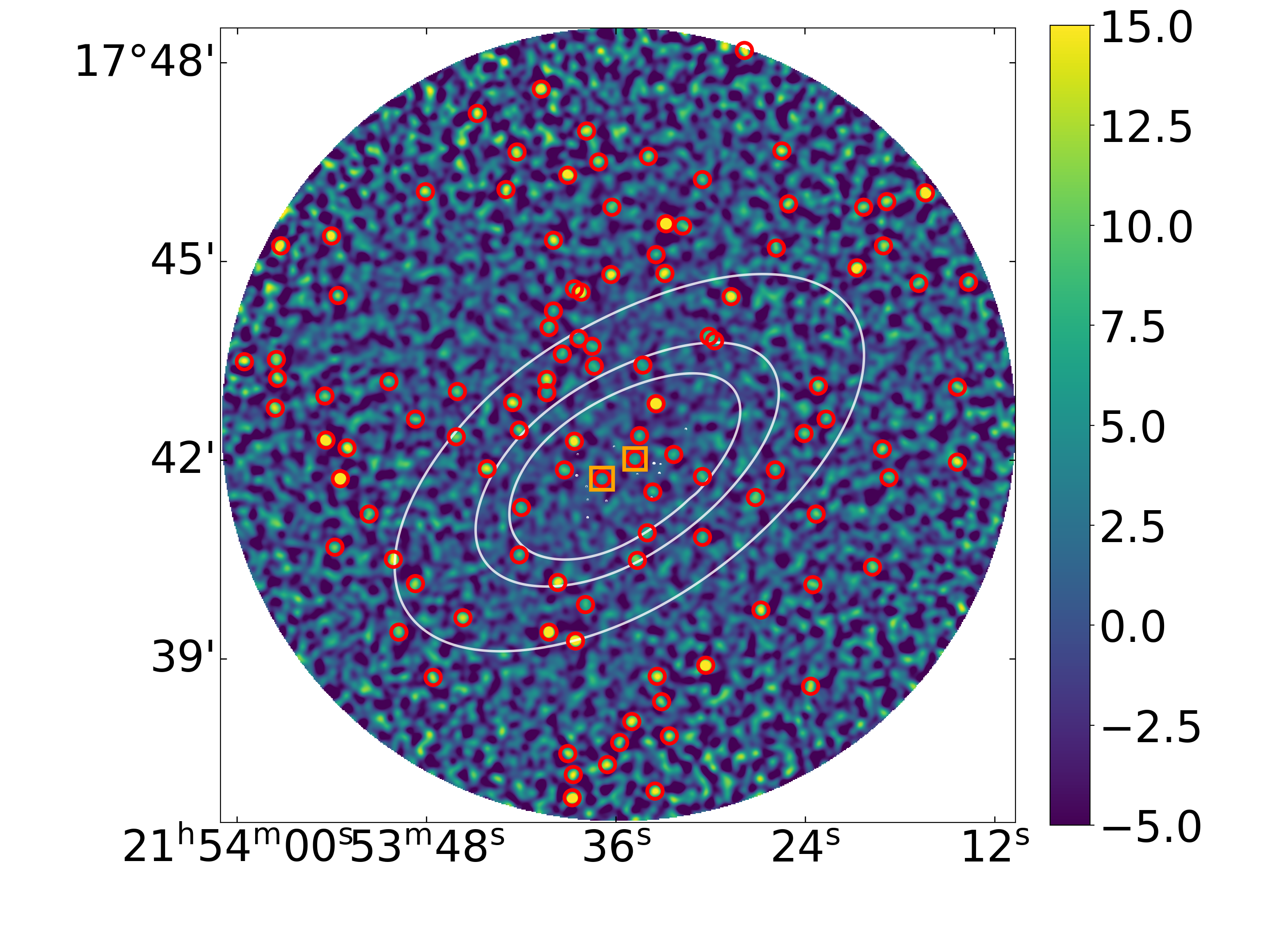}

         \label{fig:A2390}
     \end{subfigure}
     \hfill
     \begin{subfigure}[ht]{0.32\textwidth}
         \centering
         \textbf{A2744}\par\medskip
         \vspace{-2mm} 
         \includegraphics[width=\textwidth]{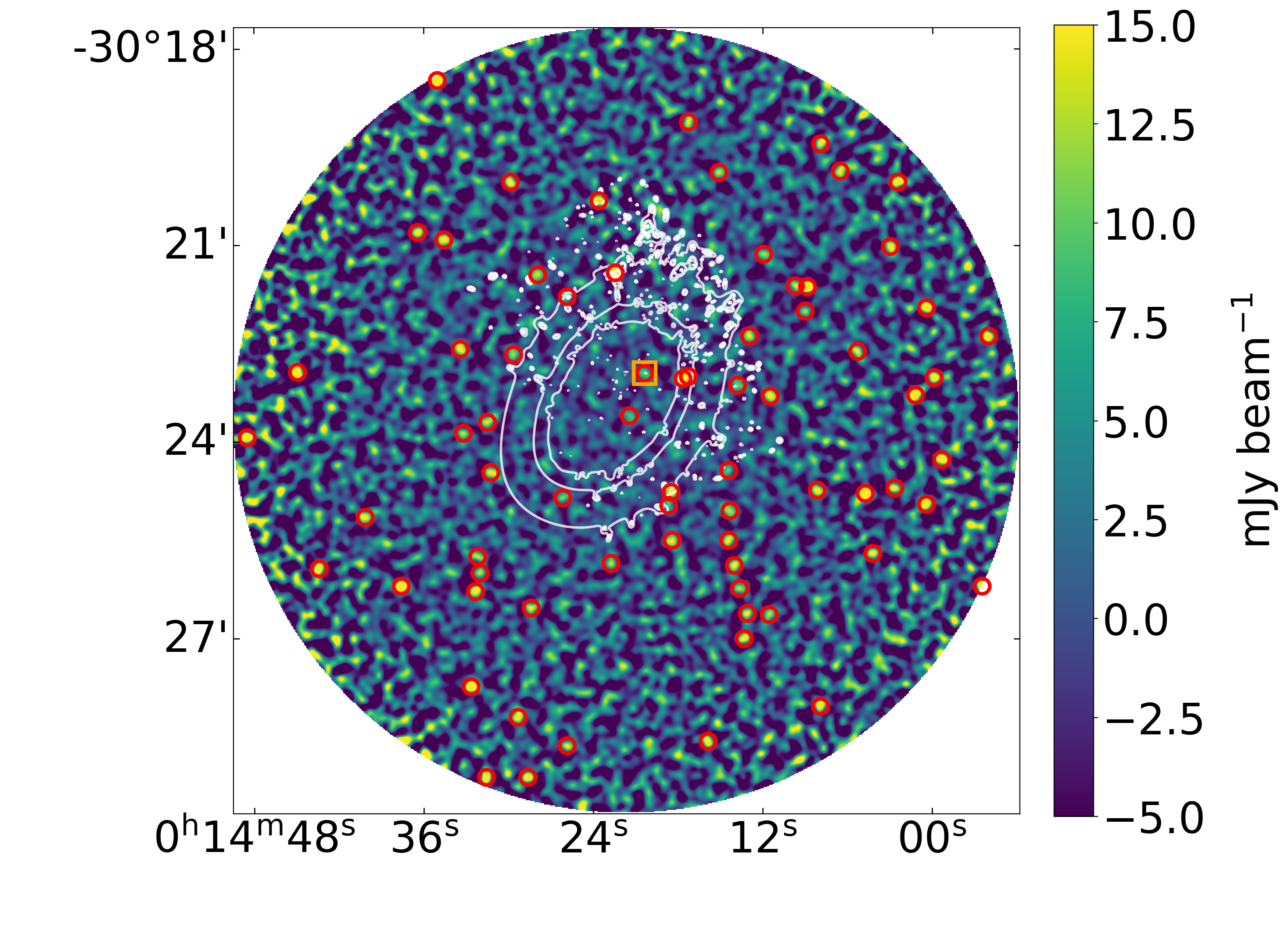}

         \label{fig:A2744}
     \end{subfigure}

     \begin{subfigure}[ht]{0.32\textwidth}
         \centering
         \textbf{A370}\par\medskip
         \vspace{-2mm}
         \includegraphics[width=\textwidth]{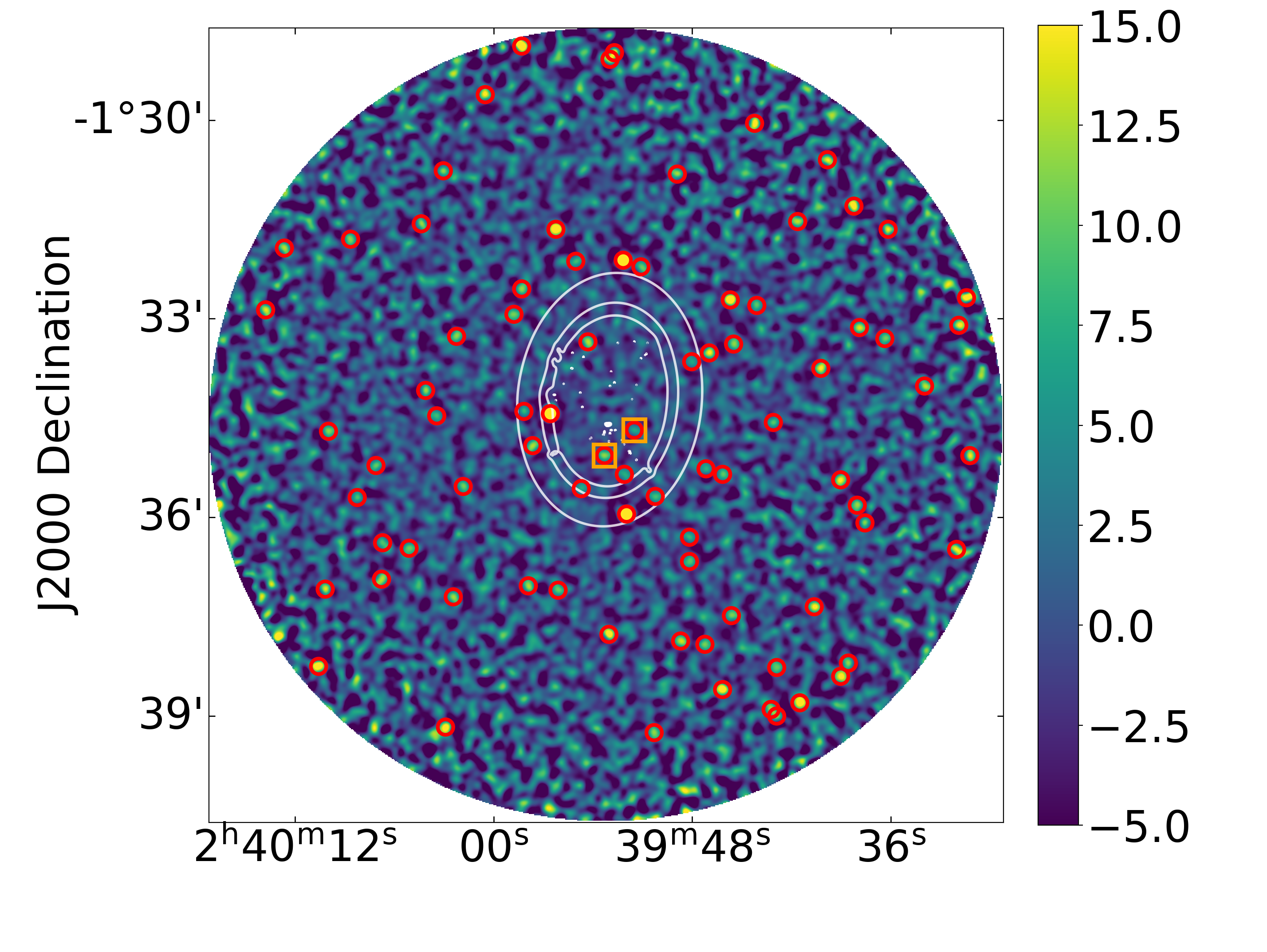}

         \label{fig:A370}
     \end{subfigure}
     \hfill
     \begin{subfigure}[ht]{0.32\textwidth}
         \centering
         \textbf{MACSJ0416}\par\medskip
         \vspace{-2mm}
         \includegraphics[width=\textwidth]{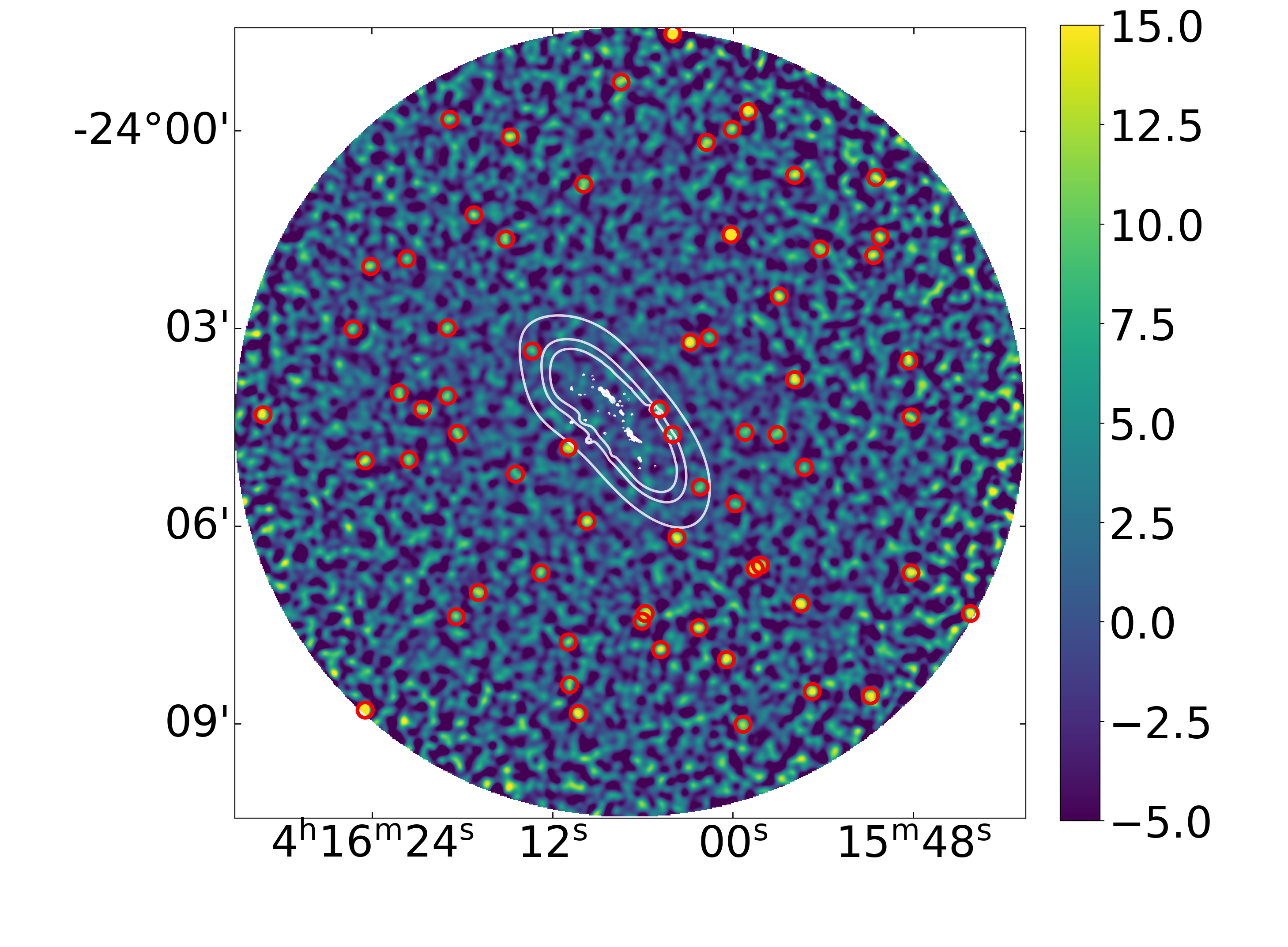}

         \label{fig:MACSJ0416}
     \end{subfigure}
     \hfill
     \begin{subfigure}[ht]{0.32\textwidth}
         \centering
         \textbf{MACSJ0717}\par\medskip
         \vspace{-2mm}
         \includegraphics[width=\textwidth]{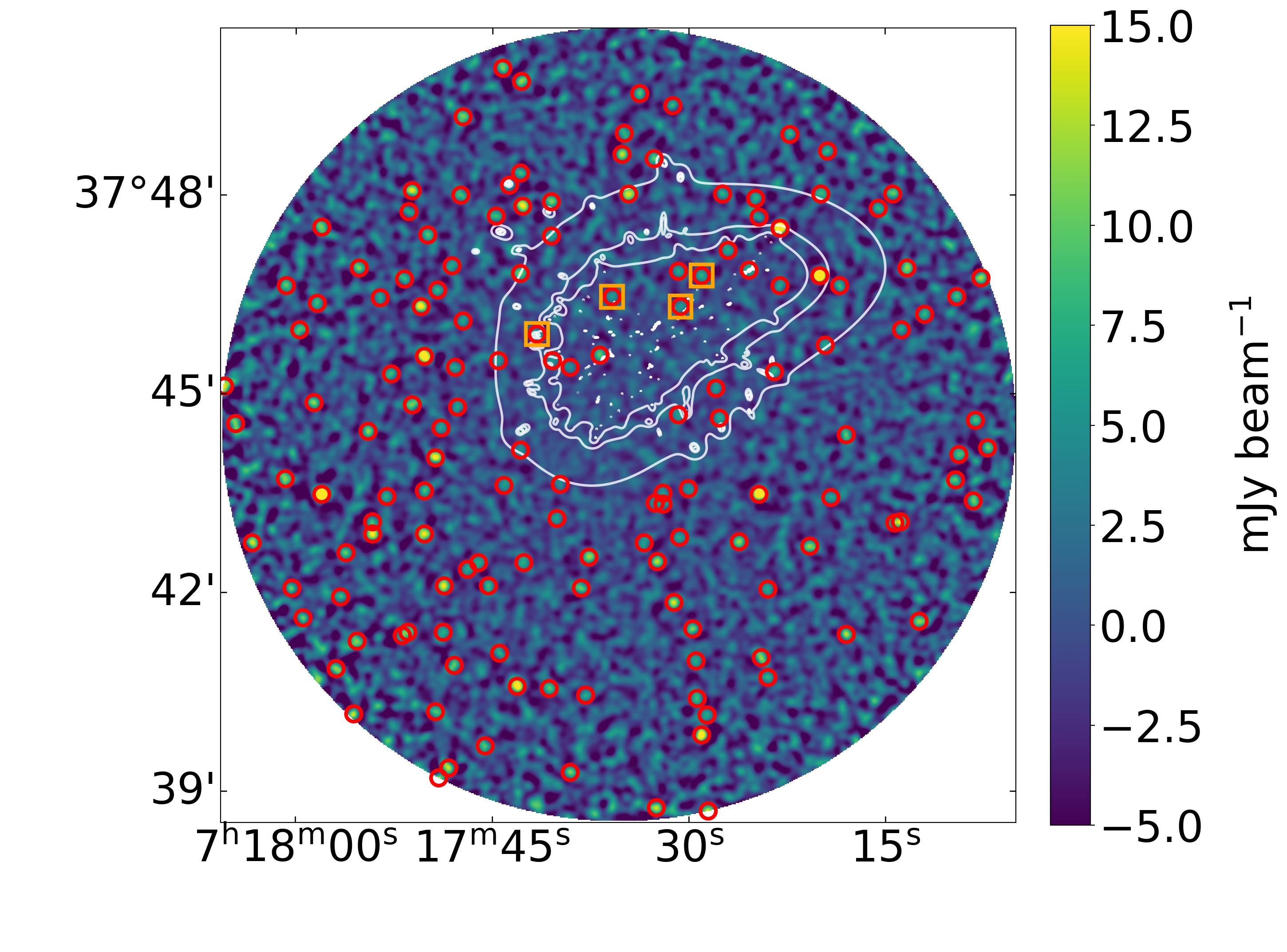}

         \label{fig:MACSJ0717}
     \end{subfigure}

     \begin{subfigure}[ht]{0.32\textwidth}
         \centering
         \textbf{MACSJ1423}\par\medskip
         \vspace{-2mm}
         \includegraphics[width=\textwidth]{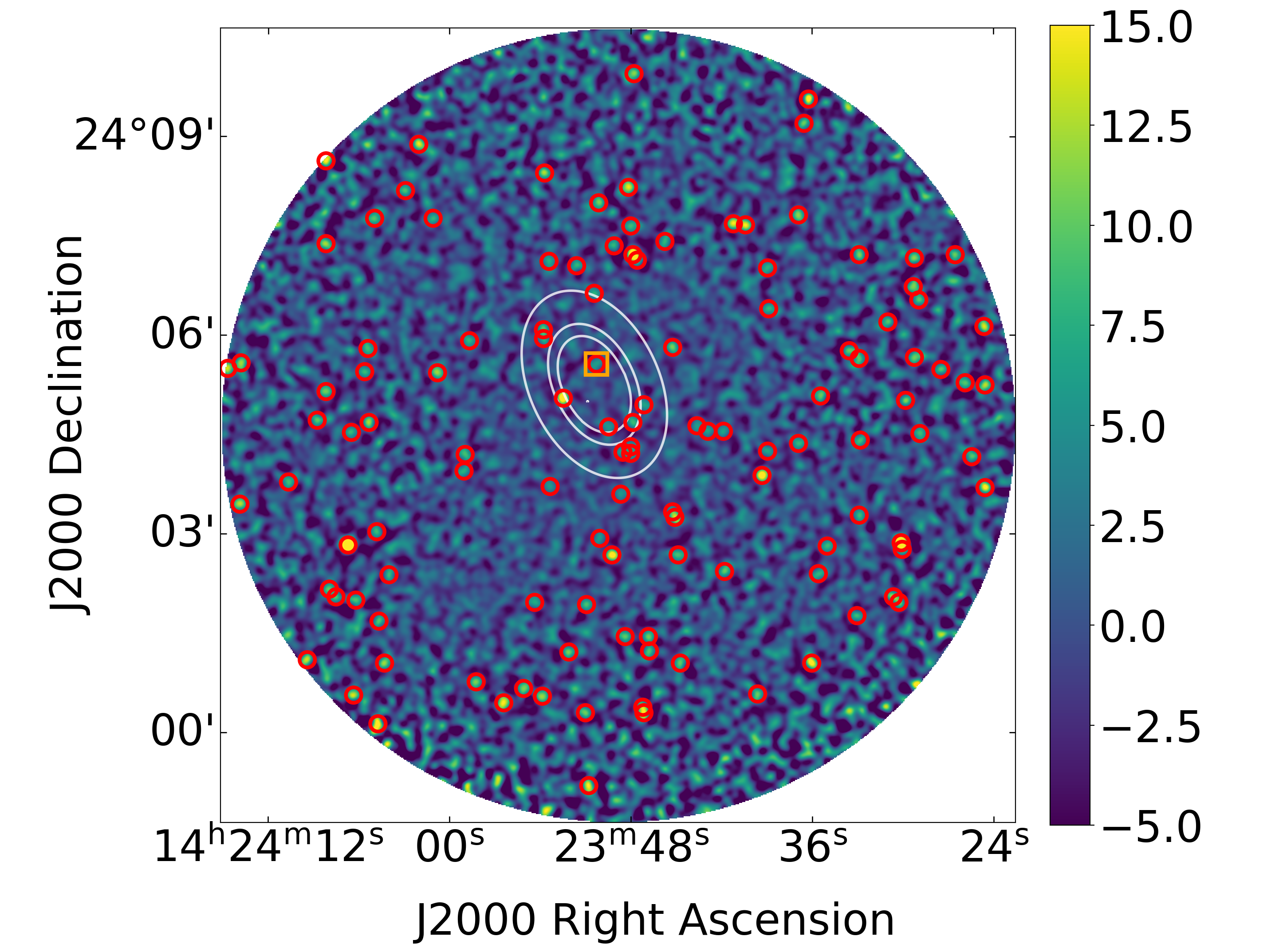}

         \label{fig:MACSJ1423}
     \end{subfigure}
     \hfill
     \begin{subfigure}[ht]{0.32\textwidth}
         \centering
         \textbf{MACSJ2129}\par\medskip
         \vspace{-2mm}
         \includegraphics[width=\textwidth]{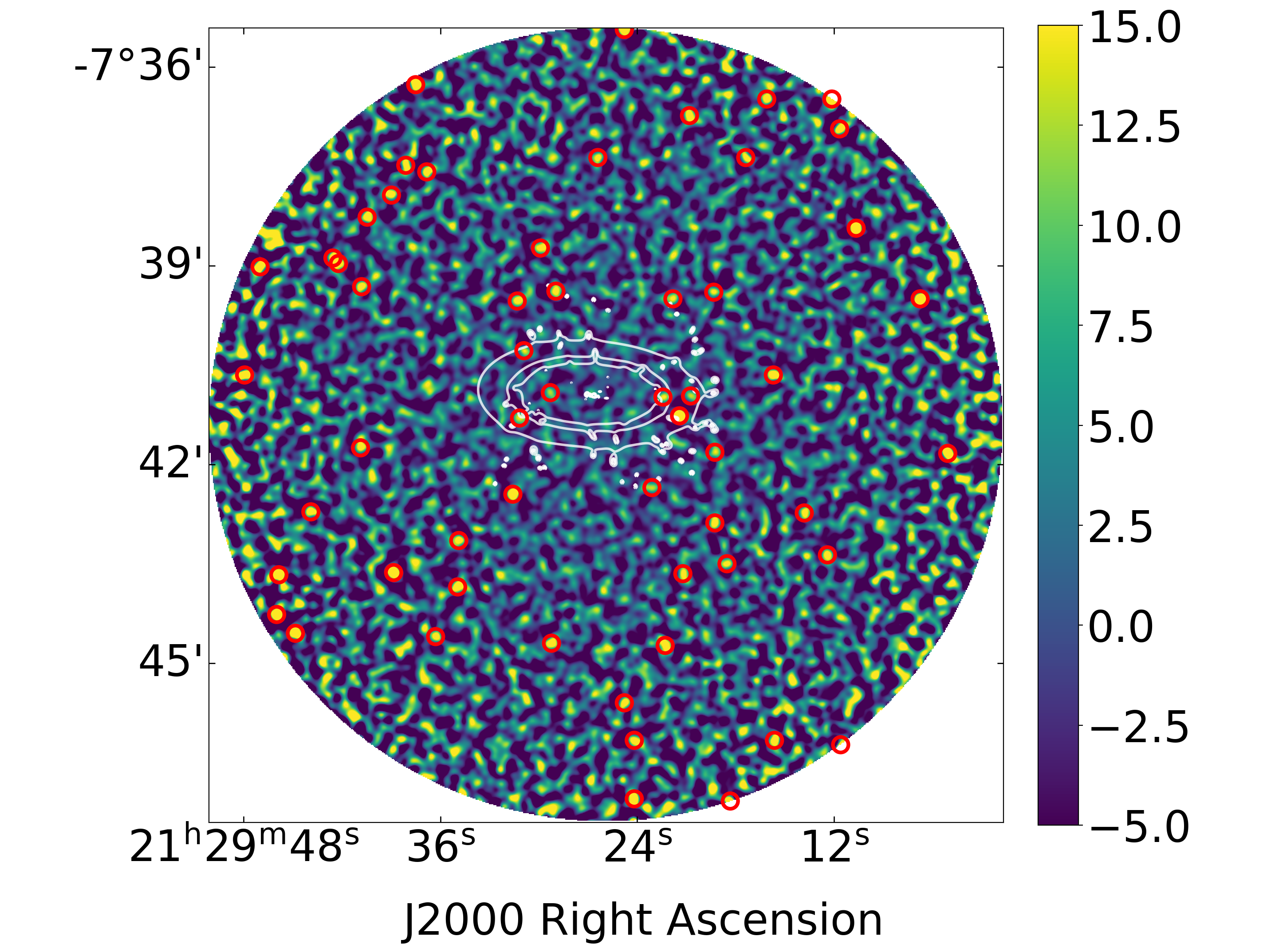}

         \label{fig:MACSJ2129}
     \end{subfigure}
     \hfill
     \begin{subfigure}[ht]{0.32\textwidth}
         \centering
         \textbf{RXJ1347}\par\medskip
         \vspace{-2mm}
         \includegraphics[width=\textwidth]{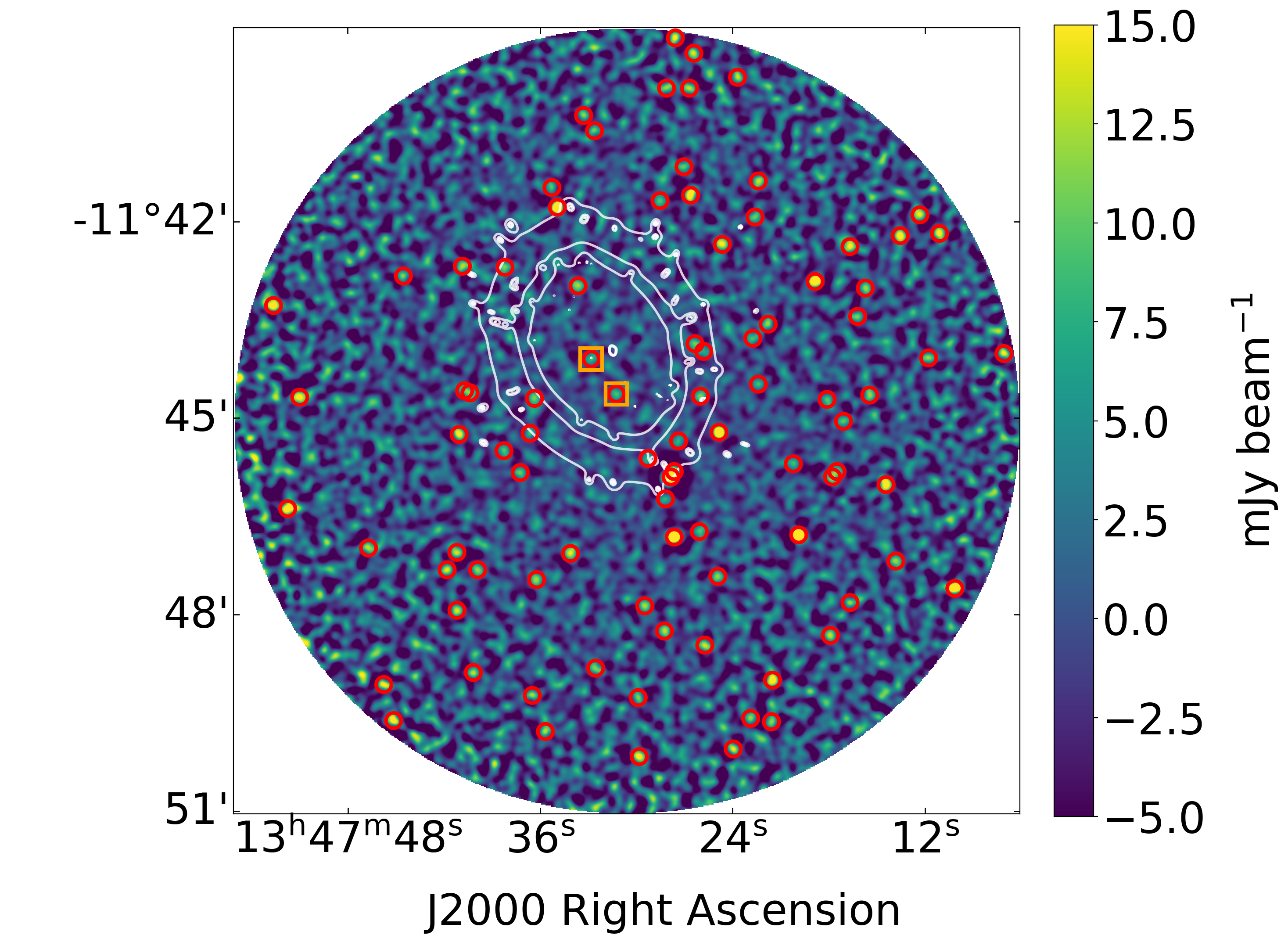}

         \label{fig:RXJ1347}
     \end{subfigure}

\end{figure}